\documentstyle[12pt,psfig]{article}
\begin{document}

{\large Testing  the
SOC hypothesis for the magnetosphere}\\

\noindent N.~W.  Watkins\footnote{Corresponding author: Fax: +44 1223 221226, Email
address: NWW@bas.ac.uk}, M. P. Freeman, \\
 British Antarctic Survey, Cambridge, CB3 0ET, UK. \\ \\
 S.~C.  Chapman, \\  Space and Astrophysics Group,\\
University of Warwick, Coventry, CV4 7AL, UK.   \\ \\
 R.~O.  Dendy,  \\ EURATOM/UKAEA Fusion Association,  Culham Science Centre,\\
 Abingdon, Oxfordshire, OX14 3DB, UK. 


\begin{abstract} 
  
As noted by Chang,   the hypothesis of Self-Organised Criticality provides a theoretical framework in
which the low dimensionality seen in magnetospheric indices can be combined with the scaling seen
in their power spectra and with observed plasma bursty bulk flows. As such, it has considerable
appeal, describing the aspects of the magnetospheric fuelling:storage:release cycle which are generic
to slowly-driven, interaction-dominated, thresholded systems  rather than unique to the
magnetosphere. In consequence, several recent numerical ``sandpile" algorithms have been used with a
view to comparison with magnetospheric observables. However, demonstration of SOC in the magnetosphere will
require further work in the definition of a set of observable properties which are the unique ``fingerprint"
of SOC. This is because, for example, a scale-free power spectrum admits  several possible explanations
other than SOC. A more subtle problem is important for
both simulations and data analysis when dealing with multiscale and 
hence broadband phenomena such
as SOC. This is that finite length systems such as the magnetosphere 
or magnetotail will by definition give information
over a small range of orders of magnitude, and so scaling will
tend to be narrowband. Here we develop a simple framework in which previous descriptions of magnetospheric dynamics
can be described and contrasted. We then review existing observations which are indicative of SOC, and
ask if they are sufficient to demonstrate it unambiguously, and if not, what new observations need
to be made?
  
\end{abstract} 
 
\section{1. Introduction: Few-parameter models for magnetospheric dynamics}

There is growing evidence that the coupled solar wind-magnetosphere
-ionosphere (SW-M-I) system, viewed as a whole, is non-equilibrium, driven, dissipative
and nonlinear (V\"{o}r\"{o}s, 1991). That this should be so is reasonable, given that the magnetosphere
is a complex system, with multiple self-interacting phenomena, occurring on a vast range of length
and time scales. A consequence of this view is that part or all of the observed magnetospheric phenomenology may be
a manifestation of physics resulting from  the underlying complexity of the whole  system.
Because of their analytical intractability, such  systems in space physics are typically studied using
 a ``large" (i.e. many-parameter) numerical simulation code.
 More recently, however, some systems of this type in nature have been shown to lend themselves
  to few-parameter descriptions (Hastings and Sugihara, 1993), which arise because, paradoxically, the complexity of the system
  gives rise to simplicity in some of its observable characteristics.
Examples of such descriptions are shown in the top row of Table 1, 
  adapted from figure 7.1 of Hastings and Sugihara, (1993).
  Starting with the simplest, ``linear plus noise" description, applied
  to the magnetosphere by Bargatze et al., (1985), we go to low dimensional nonlinear ``chaotic" models such as
  Baker et al. (1990)'s modified ``dripping tap". We then see  fractional Brownian
  motion (fBm), used by Takalo et al., (1994), as a null hypothesis against which chaos
   could be tested  and finally we have Self-Organised Criticality (SOC), the magnetospheric application
  of which is due to Chang, (1992).
 Few-parameter models of intrinsically complex systems
  have already demonstrated their value in space physics by their ability 
 to describe reproducible aspects  of the magnetosphere's behaviour 
 and   to motivate nonlinear predictive filters for geomagnetic activity
  (see the reviews of Klimas et al., 1996 and Sharma, 1995). The extent to which  such models
are applicable  bears directly on the extent to which magnetospheric (and other laboratory
  or astrophysical macroscopic plasma systems) may have predictable phenomenology.
 In consequence the study of few parameter models for energetically open but spatially confined plasma 
 systems  is a highly topical subject both with respect to the magnetospheric confinement system
  (e.g. Angelopoulos et al., 1999; Baker et al., 1999; Horton et al., 1999) and to magnetically confined 
  laboratory plasmas (e.g. Dendy and Helander, 1997; Carreras et al., 1999; Pedrosa
  et al., 1999). One possible new approach of considerable
  current interest is the SOC paradigm  introduced
  by Bak et al. (1987).

 There is a natural hierarchy of few-parameter descriptions, ordered by the extent to which
 the many coupled degrees of freedom of the system  manifest themselves.
 Broadly speaking, as we go from left to right along  Table 1, we move down the hierarchy
 of description and the large number of degrees of freedom become increasingly
 explicit in the description. In consequence the importance of an underlying theory to define the model fully
 tends to also increase. To make effective use of the theory-model correspondence in such a  table, however, 
theories must be falsifiable, as otherwise the parameters of the simple model may simply be ``tuned" to bring it into
closer and closer agreement with data. This may be a two-way process, as for example, casting  
 a given model in falsifiable form by defining which phenomena
 must be tested for helps to clarify the underlying theory.
 
 To make these abstract points more concrete, consider  Table 1. The first row shows some notable examples of 
 various simple models that have been applied  both to  the complex magnetospheric system, and
to   other such complex systems. 
The first column shows various  
properties that these models have which could be  tested for in data, provided that
suitable variables  are measurable.  If a given property can
be shown  not to be present in data then we can eliminate  models which depend on it from consideration.
   In this paper we first describe the construction
   of Table 1 by describing the four levels of description which it encapsulates. As models
   based on the SOC hypothesis  are of current
   interest for the SW-M-I coupling problem, we then
   specifically address the tests necessary to  cast SOC models in falsifiable form.
   
 
\subsection {2.1 Linear models with optional noise term}
 
 Column 1 of Table 1 is the ``linear + noise"  model, typically a linear
 differential equation with
  optionally a stochastic noise term $\Delta {\bf w}(t)$,  (adopting 
  the notation of Hastings and Sugihara (1993))
 to which we may also add  a driving term ${\bf F(t)}$:
 \begin{equation}
 \frac{d {\bf x}(t)}{dt} ={\bf g}({\bf x},t)  + \Delta {\bf w}(t) + {\bf F}(t)
 \end{equation}
 where ${\bf g}({\bf x},t)$ can only be linear in the variables ${\bf x}$.
  Physically an input-output system is linear if the  
 form of a system's response closely resembles that of  the forcing terms.
 This 
 was the first level of approximation used in the
 input-output analysis of the SW-M-I system (Bargatze et al., 1985). 
 The second and third rows of the table, labelled ``Short-" and ``long-term predictable" refer to the fact that 
in the absence of noise  ($\Delta {\bf w}(t) = 0$) the short-and long-term behaviour of equation
 (1) is  completely deterministic, while  even if an additive  stochastic
 noise term is
 present, closely-spaced initial conditions  do not show exponential divergence.   Such systems
  typically show relatively narrow-band spectral behaviour
  if the ${\bf g}$ term is dominant i.e. characteristic frequencies,
  and so we have ``no" in the  ``global scaling" row
  for this model (row 4, column 2)  to indicate that they would then
   not be scale free
  across the whole frequency range.  They may, however,
  show  regions of scale free behaviour in their frequency spectra, indicated in the table by the 
  ``sometimes" in the ``scaling regions" row (column 2, row 5). The entry ``no"
 for  ``low G-P dimension" (sixth row, column two), refers to the fact that such a system will usually appear high dimensional to the
   Grassberger-Procaccia (GP) algorithm (Grassberger and Procaccia, 1983), because of  many degrees
   of freedom of the noise term.

 \subsection{2.2 Nonlinear deterministic models}
 
 Bargatze et al. (1985) confirmed the presence of nonlinearity in the  $AE$ family of indices, 
 ($AE, AU$, and $AL$) and hence the need for a next level of approximation.
  A prototype for differential equation models which exhibit nonlinear but deterministic dynamics (see the reviews
 of Sharma (1995) and Klimas et al. (1996)) is
  the ``dripping faucet" of Shaw  (1984), which was 
  adapted to the magnetospheric problem by Baker at al. (1990).
 These models are of the form
 \begin{equation}
  \frac{d {\bf x}(t)}{dt} ={\bf g}({\bf x},t)) + {\bf F}(t)
 \end{equation}
 where unlike equation (1) the term ${\bf g}({\bf x},t)$  now contains nonlinear terms.
In the hierarchy of Table 1, this is a nonlinear  model, which can
exhibit  low-dimensional, chaotic  dynamics 
(column 3). Familiar examples of such systems in nonlinear physics include the (continuous) driven nonlinear 
pendulum and (discrete) logistic map (see e.g. Rowlands, 1990). 
In the magnetosphere this description was  inspired by an analogy between a 
dripping tap and  plasmoid ejection during substorms. The analogy was developed into a simplified magnetospheric
 model by estimating the large-scale
electrical properties of the M-I system and combining these electrical components into a 
driven nonlinear oscillator circuit model
(Klimas et al., 1992). It has been further developed into a plasma physics model by Horton and Doxas (1996).

In the case of a dissipative, driven, autonomous low dimensional system such
 as the Lorenz model, the dynamics, rather than exploring all of phase space 
 ergodically, collapses onto a low dimensional region called an {\em attractor}. This attractor has 
 fractional dimension (i.e. it is a strange attractor, in contrast for example 
 with the 2D  ellipse described in phase space by a simple linear 1D pendulum). A time series drawn from such
a system will thus  also have low  fractional dimension when tested with the Grassberger-Procaccia algorithm,
 so we write ``yes" against ``Low G-P dimension" in  column 3, row 6 of table 1. A
 strange attractor has the property that closely-spaced
 trajectories, with initial conditions  identical to within
 measurement error, will diverge strongly if they traverse  certain regions of the attractor
 i.e. repulsive fixed points (see figure 1 of Palmer, (1993) for a clear illustration of this). We thus write ``no" 
 against ``long-term predictability" (column 3, row 3), because
 in this sense, measured by a positive Lyapunov
 exponent (e.g. Rowlands, 1990), it is now not present. The significance of this 
 ``new" low-dimensional, deterministic, chaos  is that sensitive dependence on initial conditions
 arises from the ${\bf g}$ term and so exists  without the presence of ``old fashioned" stochasticity i.e. we need no 
 $\Delta {\bf w}$ term.  
Such a model can generate wide-band, scale free ``1/f" spectral behaviour when  
  near a tangent bifurcation leading to intermittency (Lichtenberg and Lieberman, 1992), 
but this  requires choice of certain values of the control parameters i.e. tuning.
 We thus write ``sometimes" for against ``global scaling" and ``scaling regions".
 ``Tuning" in  this sense
 has been considered  a weakness in the applicability of low dimensional chaos to any
 complex natural system (Bak, 1997), not only the magnetosphere.
A second practical question with such methods is that  because the model definition usually starts from
 the observables, the map  to which one applies  nonlinear dynamics must  be derived from data rather than given
 {\em a priori} from theory. One might however see this as a strength, and in practice this is  addressed by an iterative
 process whereby the parameters suggested by observation and theory are being brought closer together
 (Klimas et al., 1996).
 
\subsection{2.3 Stochastic descriptive models}

  Osborne and Provenzale (1989) showed that time series taken from certain random ``coloured 
 noise" processes, when tested with the G-P algorithm, would exhibit low dimensionality,
  and thus behave in this respect as a low dimensional chaotic system would. 
This  led to the application by Takalo et al., (1994 and references therein)
 of a third type of model,   fractional Brownian motion 
 denoted by fBm in Table 1 (e.g. Malamud
 and Turcotte, 1999), as a hypothesis
  against which to test the low dimensional nonlinear models described by the previous column.
  The suggestion of fBm  recognised the possibility that the apparent low dimensionality and fractality of the magnetospheric 
  indices was the consequence  of their being the output of an otherwise intrinsically many-degree of freedom stochastic system,
 identified by a particular ``coloured noise" power spectrum 
 (hence ``sometimes" against ``Low G-P dimension" row 6, column 4). Effectively the model is:
 
 \begin{equation}
 \frac{d {\bf x}(t)}{dt} = \Delta {\bf w}(t) 
 \end{equation}
 
A simple example is Brownian motion where the time evolution is discrete, and each step 
($\delta {\bf x} = \delta t \ \Delta {\bf w}$ )
is drawn from a white Gaussian distribution (e.g. Malamud and Turcotte, 1999). 
We then find that neither short term nor long term prediction
is possible because each step is entirely stochastic, giving us ``no" 
against ``short-" and ``long-term predictable" (rows 2
and 3 of column 3). We  note however that closely-spaced initial conditions
diverge algebraically rather than exponentially, i.e. the 
impossibility of long term forecasting here arises from the external 
stochasticity in $\Delta {\bf w}(t)$ rather than intrinsic chaos from ${\bf g}$.
Global scaling (row 4, column 4) must arise, irrespective 
of any free parameters, because there is no time scale in such a model. More complex time evolutions, 
where successive steps are taken from a fractional
Gaussian noise, are called fractal Brownian motions. A subset of such motions 
has been shown to have low G-P dimension (Osborne and Provenzale, 1989). We note that 
the presence of global  scaling or scaling regions  in the power spectra drawn from time series, 
or low G-P dimension, cannot distinguish between nonlinear
low dimensional models (column 3) and
fBm (column 4), because they are shared properties. 
The differences will only be unambiguous when one notes the different physical origins of the low G-P dimension 
between chaos and coloured noise, see Takalo et al., (1994), or when one uses 
another discriminator such as short term predictability.
 
\subsection{2.4 Sandpile models of self organised criticality}

 The most recently introduced  class of models (column 5) in Table 1 are those motivated
 by the  hypothesis of  Self-Organised Criticality (Chang, 1992; 1999, see also
  V\"{o}r\"{o}s, (1991); Chen and Holland, (1993); Robinson, (1993)). 
  SOC was first identified in (Bak et al., 1987),
   and can be modelled by,  numerical cellular automaton
   ``sandpile  models" (Katz, 1986; Bak et al., 1987)
     These discrete-variable models (Consolini, 1997;  Uritsky and Pudovkin, 1998)
     and the closely-related continuous-variable discrete-space time models (Chapman et al, 1998;2000,
      Takalo et al., 1999a;1999b; Watkins et al., 1999b)   are currently being studied for their possible magnetospheric application.
Consideration of SOC in our context was  motivated 
  in particular by the fact that SOC can account for known
 magnetospheric phenomenology such as low dimensionality (Chang, 1992) and scale free power spectra, while 
     providing a   framework for understanding  observed properties of the
     magnetosphere such as bursty transport in the tail (c.f. ``Bursty bulk flows" (Angelopoulos, 1996)).
     It may be that the long term value of SOC to plasma physics will be as a starting point     
     for more realistic ``avalanche" models of 
     turbulent transport (see Dendy and Helander, 1997, for
     more on this question, as applied to laboratory plasmas). However, at this stage of its consideration
     with respect to understanding
     in magnetospheric  physics it remains useful to consider Bak et al.'s original, sandpile model-based, definition of SOC
     in the framework of our table of observables, as it is used explicitly
     or implicitly by much of the  current work  on SOC in this and other fields.

Bak et al. (1987, henceforth BTW87) originally proposed SOC to
explain  the apparent ubiquity of both spatial fractals  and
``1/f" spectra in nature.  
They observed it in a class of numerical 
cellular automata, called ``sandpile models" 
for which analogous  continuous thresholded diffusion  
equations have since been shown to  exist (Lu, 1995).
The equations are modified from stochastic diffusion equations (Pelletier and
Turcotte, 1999) which have a form such as
\begin{equation}
\frac {\partial {\bf g}({\bf x},t)}{\partial t} = \nabla^2 {\bf g}({\bf x},t) + \Delta w(x,t)
\end{equation}
by the introduction of a thresholding process (Jensen, 1998), e. g.
\begin{equation}
\frac {\partial {\bf g}({\bf x},t)}{\partial t} = \nabla^2  
{\bf g}({\bf x},t) \Theta ({\bf g}({\bf x},t) - {\bf g_c}) + \Delta w(x,t)
\end{equation}
where the step function $\Theta$ initiates diffusive transport when the variable ${\bf g}$ reaches a critical
gradient ${\bf g_c}$.
The main debate centres on how to motivate and satisfactorily introduce this {\it ad hoc} thresholding term 
(see Lu, 1995;Jensen, 1998) but several properties, such as the low frequency 
power spectrum  may be
dependent only on equation (4) and the nature of the boundary terms (Jensen, 1998), 
and so may be common to both SOC and stochastic diffusion.

The behaviour  classified by BTW87 as SOC is the evolution of
the medium described by the cellular automaton or 
differential equation models  from arbitrary initial conditions to a non-equilibrium 
but steady state, ``self-organisation". The medium  then evolves by dissipating
energy on all scales via thresholded reconfiguration/energy 
release events called ``avalanches". 

The assertion by Bak et al. (1987;1988) that the 
observed scale free, and hence power law,  distribution of the
size of these energy release events (the ``avalanche distribution") measured
 the arbitrary response of a self-organised fractal structure in the medium to perturbations introduced
by random fuelling was the reason for their  use of the term ``critical". Their
analogy was with the scale free critical state 
associated with phase transitions in critical phenomena (Huang, 1987). The common
  observable features cited by Bak et al. were that both systems 
  were globally scale free (hence we may write ``yes" in column 5, row 4 of table 1),
  and also that a finite size scaling analysis gave a good data collapse, as it would in
  a {\it bona fide} critical system (Cardy,1996).
   The combination of a scale free response to perturbations with the release of 
this energy by random unloading events, was expected to give rise to a power law
frequency spectrum. This was expected to be ``1/f" and hence to explain
the ubiquity in nature of noise with  correlations on all time-scales. Unlike
a ``$1/f^2$" spectrum  above a characteristic frequency which can 
be explained in many systems simply
by random switching of levels, the appearance of ``1/f" ($f^{-\beta}$)
spectra where the spectral index is between about 0.8 and 1.4 is a 
long standing problem in many branches of physics (Jensen, 1998).

In summary, in this
 picture, the SOC hypothesis would  be that: ``extended driven systems will tend to self-organise to 
 fractal structures which dissipate energy on all scales in space and time, and hence give rise to
scale-free ``avalanche" energy burst distributions and ``1/f" noise".
More recent sandpile algorithms which allow fuelling to continue while unloading occurs
have a broken power law spectrum and so we have added ``yes" to column 5, row 5
as well (see also section 3.1.2). 
 
SOC behaviour, as diagnosed by scale-free energy release and/or ``1/f" 
power spectra, has since been claimed for many systems in nature (see chapter 3 of
 Jensen (1998) for a compact review, and Rodriguez-Itube and Rinaldo (1997) for a longer exposition in the 
particular context of fractal river networks).  
At this point, we simply note that a definition of SOC in terms of what an SOC 
system does can only be used to identify an SOC system if no other system does
exactly the same things. Identification of global scaling, shared by SOC, fractal Brownian motion
and low dimensional chaotic systems when intermittent is, for example, thus
not an unambiguous test. It is for this reason that we 
have used  Table 1, as a guide to how the ``footprint" of SOC may be
more unambiguously defined.   We have left the other rows as question marks 
because BTW87's sandpile model definition of SOC was not unambiguous in these respects,
and these issues are still under study. 
The wider definition of SOC used by Chang, (1992), predicts low dimensional behaviour 
i.e. ``yes" in column 5, row  6, at least close to criticality, while many workers
have taken the predictions for column 5, rows 1 and 2 to be ``no" e.g. the remarks
of Consolini, (1997): ``In fact, if the magnetospheric dynamics could be the
result of a low-dimensional dimensional chaotic dynamics, we could have some 
hope to forecast the evolution. On the contrary, the existence of a critical
state removes this possibility, because the fluctuations of the
system at a critical point are completely unpredictable". This
is equivalent behaviour to fBm. See also Bak (1994) on this point
where sandpile models are asserted not to show sensitive dependence on initial conditions
i.e. they are unpredictable on both long and short timescales but not chaotic. 
The similarity of SOC to fBm with regard to the phenomena in Table 1 might
suggest that SOC adds nothing to fBm. However, there are differences. One is
the fact that an SOC system releases energy by means of avalanches, 
effectively a new observable property, which we have  thus 
indicated by adding a  row in Table 1 to those used by 
Hastings and Sugihara (1993). We are indebted to  a referee for the 
suggestion that avalanche models may also have different phase
spectra to the (usually random) phase behaviour of noise.
A second advantage   is that SOC can be explained in terms of an underlying theory and
encapsulated in terms of sandpile models, which begin to allow explanation in
terms of the underlying plasma physics of the system. A third advantage is that the release of energy
by avalanches is suggestive both of bursty transport in plasma confinement 
systems (e.g Carreras et al., 1999;Pedrosa et al., 1999), and, possibly, the substorm problem (Chang, 1992; Consolini, 1997).

The study of SOC in solar terrestrial-physics has proceeded initially
through comparison of signatures
in data, particularly the AE/AU/AL indices, 
with analogous signatures in ``sandpile model" realisations
 of SOC. However, as we will now show, these signatures are not all unique to SOC, and
  the combinations in which they appear may be model dependent. 
 Furthermore, we recall that the original proposal of the relevance of SOC to 
 the SW-M-I system (Chang, 1992) was not predicated exclusively on a definition of SOC
 derived from sandpile models. To minimise possible confusion in this
 rapidly developing area, two questions are  addressed.
These are i) which experimental signatures will be needed to  distinguish
  unambiguously between SOC and, for example, deterministic chaos
  and ii) what are the predictions of SOC models which will
  be robust against fluctuations in the input, or limited 
  station or satellite coverage etc?

\section{3 Towards unambiguous tests of SOC}

Having decided what the predictions of SOC are which may be confidently entered in Table 1, we now go on to see what the observations currently available enable us to 
say. The question immediately arises as to whether, Picture A), the SOC system is seen as being the 
complete magnetosphere (``global SOC"), in which case ${\bf x}$ in equation (5) are the system state variables, for which the
AE indices (Davis and Sugiura, 1966) have been taken as proxies; or, Picture B),  SOC is more local in scope (``local SOC"), and 
plays a role in generating, stabilising and destabilising the magnetotail current sheet,
in which case ${\bf x}$ might be a locally-measured magnetic field or the field seen
in an MHD-derived sandpile simulation. Picture A) is closer to that given in  Uritsky and Pudovkin (1998) and
 Consolini (1997; 1999), while Picture B) seems to us to be 
  the motivation for Takalo et al., (1999a;b;c). Because any approximation will have a natural maximum
scale of applicability, the idea of ``local SOC" is not the contradiction it may at first
appear to be. 

It has been objected that if A) were true, all system-level outputs should
show global scaling and that some are observed 
not to have this property (Borovsky and Nemzek, 1994). However Chapman et al. (1998), used the 1-dimensional
avalanche  model of Dendy and Helander (1998), to illustrate a system in which the internal
energy release showed scaling while energy flowing out of the system (``systemwide")
did not, a feature seen in some  other sandpile models (e.g. Pinho and Andrade, 1998). Until pictures A
and B can either be distinguished or reconciled, care must be taken not to justify one
using measurements consistent with the other and vice versa. We thus first (section 3.1) examine those 
system level outputs in which evidence of SOC has been claimed, and then (section 3.2) briefly consider
evidence for SOC on more internal scales.

\subsection{3.1 Remote Observations of system outputs}

So far the main global dataset  for testing for SOC has been the $AE$ indices. This is because, since Bargatze
et al. (1985), a candidate dynamical variable for all the models discussed above has been the energy 
dissipated by the magnetosphere into the ionosphere, for which most workers have taken  
 the Auroral Electrojet Index ($AE$) to be a proxy.

\subsubsection{3.1.1 Global scaling: Power law power spectra}

  The work of Tsurutani et al. (1990) described the power spectrum of $AE$ as  ``broken power law",
 in that the high frequency behaviour is
 approximately $1/f^2$ while the lower frequency behaviour
 is approximately $1/f$, with a break at (1/5) hours$^{-1}$.
This  ``1/f" behaviour has been cited as evidence of SOC 
in the magnetosphere (Consolini, 1997; 1999; Uritsky and Pudovkin, 1998). Two
main scenarios have been advanced, in one the ``1/f" spectrum is seen as arising from interactions between
correlated avalanches, which would then be interpreted as substorms (Consolini, 1997); while in the second
the ``1/f" behaviour (Uritsky and Pudovkin, 1998) is related in part to the input,
which  is allowed to modulate the threshold values in the sandpile algorithm.

The  first apparent complication in this interpretation 
is that the $AE$ spectrum is ``broken" i.e.  a ``1/f$^2$" high frequency part has been reported,
 whereas criticality in BTW87's original picture was expected to give   
 long-period correlations and hence a ``1/f" spectrum (Jensen, 1998). The resolution, due to Consolini (1997),
is discussed in the  section 3.1.2. The second,  more fundamental,
problem is that a ``1/f" spectrum in the ouput of a system could only be an unambiguous indicator 
of SOC if this spectrum is not being passed through from the input. The fact that the input spectrum
of the solar wind follows $AE$ closely over  the low frequency ``1/f" range that 
concerns us here (Tsurutani et al., 1990; Freeman et al., 1998) suggests that the power spectrum 
should not be used for this purpose. In this context, the high degree of predictability of $AE$ from the
solar wind input is suggestive (e.g. Baker et al, 1997), as is the fact that the
power spectrum of the signal from a neural
network prediction of $AE$ has ``1/f" form (Takalo et al., 1996). We return to this question in section 3.1.3 
when we discuss the avalanche statistics. The possible ability of some avalanche
 algorithms to emulate a nonlinear filter, and show sensitivity to the distribution of the input fuelling rate 
(Takalo et al., 1999c); or conversely to eliminate traces of fluctuating input
 (Watkins et al., 1999b), increases the relevance of this question.

\subsection{3.1.2 Scaling Regions: Spectral breaks}

If  the system is known to be  SOC  {\it a priori} or from other tests, the presence of a high frequency ``1/f$^2$" component is
 understandable. This is because  recent work (notably Hwa and Kardar, 1992) has shown
that a ``running" sandpile (and hence SOC differential equation models such as the example  used by
(Takalo et al., 1999a)) can show this type of ``broken" power spectrum. The reason is that  allowing the 
fuelling to occur on a similar time scale to the unloading events  permits a bursty 
``1/f$^2$" power spectrum of individual avalanches to co-exist with the ``1/f" power spectrum 
which is ascribed to interactions between events (see Jensen, 1998). If the bursts are identified with substorms then the 
break at 5 hours will be related to the maximum duration of a substorm.
Furthermore, the original Bak et al. (i.e. ``non-running") sandpile model
  was quickly  shown (Jensen et al., 1989)  to 
  produce a ``1/f$^2$" spectrum in its energy release events, illustrating  
  that the although the pile is in critical state,
  shown in particular by the finite size scaling of the avalanche
  distribution (Bak et al., 1988), the critical state  is not revealed by the energy 
  release power spectrum. 
  
The complications in this very appealing simple interpretation 
arise for two main reasons. One is that we do not know {\em a priori} that the
system is SOC, so mapping an output variable of the sandpile model to the observed $AE$ spectrum is not a unique
process. An alternative way to get a broken spectrum of the form shown by 
Tsurutani et al. (1990) for $AE$ is in
 a boundary driven 2D sandpile of the BTW87 type (see figure 4.6 of Jensen, 1998).
    In this   case the variable whose spectrum is obtained is not the transport
 of sand over the edge of the pile but the sum of the dynamical
 variable ${\bf g}$ across the pile i.e
 \begin{equation}
 <g(t)> = \sum_{i,j} g_{ij} \equiv \int d {\bf x} \ {\bf g}({\bf x},t)
 \end{equation}
  As with Tsurutani et al, 1990, the spectrum shown by Jensen (1998) is
 $\sim 1/f$ below a critical frequency and $\sim 1/f^2$ above, with
 the break set by a time scale $T_{max} (L)$ corresponding to
 the longest avalanche possible in the system. This would imply that
 $T_{max}$ is related to the system size $L$, and furnish a possible test if the system's
 value of $L$ could be varied significantly. In other words, the robust property is the break
 itself rather  than the variable whose broken spectrum is being calculated.
  
The second problem is that the broken  power law spectrum for $AE$ cannot be uniquely interpreted as the
output of an SOC system because other types of physical system can produce
power spectra which  show global scaling or scaling
over a region or regions.  The example of global scaling, i.e. 
scaling over a very wide bandwidth, discussed in section  2.3
was  simple Brownian motion which has an $f^{-2}$ spectrum at all values
of $f$. A less well known example of  scaling over a restricted region is the ``random telegraph", a random sequence of square pulses (i.e. states +1 or -1) 
switched at Poisson distributed intervals which gives $f^{-2}$ for frequencies higher than the inverse
correlation time but has a flat spectrum (because uncorrelated) for lower frequencies (Bendat, 1958; Jensen, 1998). 
It is very  important to note that the ``1/f$^2$" part of the spectrum here is due entirely
to the exponential autocorrelation of the pulses, and is not the same as  the intrinsically
scale free, and wideband,  behaviour of a coloured noise source such as Brownian motion. If the  lifetimes of the correlated
pulses extend over two orders of magnitude in time then so will the ``1/f$^2$" spectrum, and 
so a test such as the second order structure function 
(see Takalo et al., 1994 and references therein), or a variance histogram (i.e. Fourier
power spectrum)  will be unable to distinguish this ``trivial" apparent
scaling from the ``interesting" scaling resulting from coloured noise. 
A similar problem whereby level changes with a $1/f^2$ spectrum might mask an intrinsic Kolmogorov spectrum
was treated for solar wind turbulence by Roberts and Goldstein (1987).

 The fact that  the high frequency scaling region in the spectrum of
 $AE$ might arise from a cause other than SOC is important in our application
because $AE$ is known {\it a priori} to be a compound index which
 mixes driven and unloading effects (Kamide and Baumjohann, 1991). This mixed origin is  reflected by its
power spectrum (Freeman et al., 1998), structure function (Takalo et al., 1994), and 
avalanche distribution (section 3.1.3 and Freeman et al., 2000). In consequence
a suitable ``null hypothesis" for the power spectrum against which the SOC models so far proposed should be 
evaluated is that $AE$ consists of a solar wind driven ``1/f" component - arising from the $DP2$ 
convection electrojet (Kamide and Baumjohann, 1991) - and a
random unloading $DP1$ 
substorm electrojet component which looks like ``1/f$^2$" over two orders of magnitude in frequency
and appears predominantly in $AL$. By analogy with section 3 we may call this Picture C (``no global 
SOC").

A possible avenue for  testing Consolini's (1997) ``interacting burst" interpretation of the ``1/f" spectrum  would then be to see if the correlation 
properties of the ``1/f" part of the power spectrum of $AE$ differ in any way from those of 
Akasofu's $\epsilon$ parameter, which estimates the componenent of solar wind  Poynting flux
entering the magnetosphere. If they do, this adds
support to the possibility that the bursts may be correlated with each other as a result of
a process which occurs in the magnetosphere itself; rather than
the ``1/f" behaviour being explained by the long-period
correlation already present in the solar wind's power spectrum.

\subsection{3.1.3 Global scaling: Avalanche distributions}

Because of the above concerns, we see a better candidate for an unambiguous indicator 
of SOC as being the
statistical distribution of  energy released by  individual ``events".  Since the work of Bak et al.
 an SOC system has been expected to show a ``power law" probability distribution for 
 this quantity.   Consolini, (1999) has plotted the distribution $D(s)$ of a burst
 measure $s = \int_{\Omega} (AE(t) - L_{AE}) dt$, formed from $AE$
 where $L_{AE}$ was a
 running quiet time background level of $70 \pm 30$ nT, and each
 integration was taken over a period $\Omega$ where the integrand was
 positive (a burst).  The $AE$ data used covered the period from 1975
 and 1978-1987.  The distribution obtained could be described  by an
 exponentially cutoff power law (Consolini, personal communication, 1998)
 extending the result previously obtained by Consolini, (1997)
for data from 1978.
 
 The presence of such a power law suggested a magnetospheric analogue of the Gutenberg-Richter
  law in seismology, and has played a significant role in generating the interest in SOC 
 in magnetospheric physics.  Both its existence and its apparent robustness with activity level
 require confirmation and explanation whether by an SOC theory or another one. Although
 both a simple power law, or the above exponentially cut off power law are possible fits,
Consolini (1999) has recently demonstrated that a better description for the burst size
distribution of $AE$ is an exponentially modified power law with a small lognormal ``bump"
component. If the system is SOC, this requires an explanation of the ``bump",
which may be found in the different behaviour of internal and systemwide
dissipation in some sandpile models (Chapman et al., 1998) or in subcritical
dynamics in the SOC system (Consolini, 1999). 

\subsection{3.1.4 Global scaling:  Lifetime distributions}

The SOC hypothesis had earlier led Takalo (1993) and Consolini (1999) to examine 
the distribution of lifetimes of the bursts. This is potentially a stronger indicator of SOC than the burst size, 
because exponentially modified power-law burst size  distributions can also be generated
 by randomly quenched, exponentially growing
instabilities in an otherwise non-critical medium (Aschwanden et al., 1998). $AE$ was found (Consolini,1999)
to show a exponentially modified  power law distribution of lifetimes, but with evidence of a ``bump" 
at around 100 minutes. The ``null hypothesis" mentioned in 3.1.2 (Picture C)
led Freeman et al., (2000) to calculate the analogous 
burst lifetime distributions for $AU$, $AL$ and
the solar wind quantities $\epsilon$ and $v B_s$. They found exponentially modified power laws
with very similar slopes for all quantities, but
the ``bump" was only found in the AL and AE, magnetospheric component. This suggests that 
the ``bump" is of intrinsically magnetospheric origin (due to the DP1
 current (Kamide and Baumjohann, 1991)) while
 the scale-free burst lifetime distribution  may actually be of solar wind origin (Freeman et al., 2000),
if the DP2 current system (Kamide and Baumjohann, 1991) 
is transparent to the driver.

\subsection{3.1.5 Other tests: Predictability, low dimensionality}

It will be necessary to examine the  sandpile models and other realisations of SOC in more detail before
we can say with certainty what they predict for the remaining rows of column 5, table 1. Bak (1994)'s arguments about
long term prediction are based on the assertion that  $\delta$, the separation of two initially infinitesimally
close trajectories in a BTW87-type sandpile model 
grows with time quadratically, $\delta = a t^2$, rather than exponentially, $\delta = e^{\lambda t}$, as would
be the case in a chaotic system such as that of column 3 (where $\lambda$ is the Lyapunov exponent). This 
assertion needs to be tested in other sandpile models and in data from candidate systems. It is a potential
discriminator between chaos and SOC.

The demonstration that an SOC system can show low dimensional behaviour was given by Chang (1992;1999) on
the basis of a more general formulation of SOC than the sandpile-inspired one of Bak et al. It thus remains
to be seen in general what sandpile models predict for dimensionality.  

\subsection{3.1.6 New tests: Intermittency and laminar time}

The open questions described in the previous section and the ambiguities necessarily present in the data
reviewed in sections 3.1.1 to 3.1.4   mean that at present it is not possible to completely eliminate any
of the models discussed in Table 1, except the artificially simple linear model which was included for completeness.
New tests are thus required. An example of such a test
is the degree of intermittency present in the time series. 
Consolini et al. (1996)  showed that the $f^{-2}$ spectral regime of $AE$ might be described as an
$f^{-1.8}$ regime corresponding to the inertial range of a turbulent system, with an exponent modified from the
Kolmogorov value by the presence of intermittency. These authors showed a good fit to the p-model of turbulence, 
also shown in the solar wind by Horbury and Balogh (1997). We are thus not presently able to distinguish between
intermittency intrinsic to $AE$ and that due to the solar wind driver. Further work on this topic
is likely to prove valuable (see also V\"{o}r\"{o}s et al., 1998).

More recently, it has been claimed (Boffetta et al., 1999) that the probability density $D(\tau)$ of 
time intervals $\tau$ between bursts (the ``laminar time") can be used to distinguish an SOC system of
the BTW87 type, which has exponential $D(\tau)$, from a shell model of turbulence,
which has power law $D(\tau)$. It might seem that the power law $D(\tau)$ for
AE shown by Consolini (1999) would rule out SOC in the global magnetosphere.
However, as emphasised by Einaudi and Velli, (1999), the predictions
for $D(\tau)$ are not in general known either for more realistic
SOC models or for all  turbulence models. The relevance of this work to the issue
of ``sympathetic flaring" (Boffetta et al., 1999 and references therein)
 in solar physics is likely to give rise to further 
exploration of this topic, and hence magnetospheric application.

\subsection{3.2 Local observations of current sheet dynamics}

It is fair to say that, for the above reasons, the evidence of SOC in the largest scale
outputs of the magnetospheric system, measured by $AE$ and $AL$, is not yet persuasive. 
The main problem is that the behaviour of the $AE$ indices is similar to that of the solar
wind in a number of respects. If an intrinsic and a solar wind component are always present,
then testing for SOC in these compound indices will always be problematic, as  will the 
interpretation of results based on them (c.f. Consolini, 1999; Freeman et al., 2000).

It may be more instructive to study  regions of the magnetosphere where the effect of
the solar wind input is less directly visible, and 
recent attention has been focused on SOC as a model of the magnetotail
current sheet. So far this has been achieved  by 
truncating the ideal MHD equations i.e. replacing the convection term in
\begin{equation}
\frac {\partial {\bf B}({\bf x},t)}{\partial t} = \nabla^2 {\bf B}({\bf x},t)  + \nabla \wedge ({\bf v} \wedge {\bf B})
\end{equation}
by a source term, resulting in an equation analogous to (4) 
\begin{equation}
\frac {\partial {\bf B}({\bf x},t)}{\partial t} = \nabla^2 {\bf B}({\bf x},t) + \Delta w(x,t)
\end{equation}
and then introducing one of several possible thresholding terms c.f. equation (5)
(Vassiliadis et al., 1998 ; Takalo et al, 1999a;b;c) to map the problem onto a cellular
automaton or a  differential equation like that of (Lu, 1995). In view of the limited applicability
of such  non-self consistent models it is encouraging that reduced MHD simulations of turbulence are also
demonstrating SOC phenomenology (Einaudi and Velli, 1999). 
 
 Consideration of SOC as a model for the magnetotail may also be motivated 
by the suggestion by  of Zelenyi et al.,(1998) that the tail exists as a critical percolation cluster.
Critical percolation, whereby an avalanche can extend exactly to the maximum scale length of a system rather than
just below or just above it, was the original proposed explanation for  the relevance of criticality
in SOC  (Bak et al., 1988). The idea has recently been further developed to explain
how  self-organisation occurs in sandpiles (Zapperi et al., 1997) in a picture whereby the edge fluctuations
drive the system back to a critical percolation state.

\section{4.  What signatures of SOC are robust enough to be detectable in  ``real world" data ?}

SOC is of particular interest to magnetospheric physics because it is
  robust, in the sense that the characteristic observed behaviour does
  not necessarily change greatly over wide ranges of parameter space.
  This robustness is thus distinct from the scale invariant phenomena
  that arise near critical points such as phase transitions and hence
  in restricted regions of parameter space (Huang, 1987).  SOC
  systems are in this way also distinct from chaotic systems such as
  the dripping tap which frequently show radically different
  types of behaviour as control parameters change, a point emphasised
  by Bak (1997, pages 29-31). They are not however, as easily distinguishable
  from fBm.
  
   However we still need to ask what
  aspect or aspects  of the magnetosphere's behaviour would be both sufficient to
  uniquely identify SOC and yet also  robust enough to be seen
  under the wide range of activity levels exhibited by the
  magnetosphere and the solar wind i.e., to pick the specific robust
  discriminator. The slowly driven condition is of particular interest in the
magnetospheric context because most SOC simulations have been
conducted in the limit where the rate of inflow is ``slow" relative to
dissipation. Watkins et al., (1999b); and Chapman et al., (1999) have recently studied the
question of  how robust 
the magnetospherically relevant aspects of SOC are to changes in the inflow 
rate. They found that the power law avalanche distribution was preserved for the largest 
values of internal energy release, and gave arguments as to why this should be so. This result
may give confidence that such a distribution, if shown on other grounds to be unique to SOC, will be
able to be used as a diagnostic. Similar studies for the power spectrum are being carried out in an MHD-derived
model by Takalo et al., (1999c).

\section{5. Conclusions}

 We have attempted to identify the distinguishing observable features of different few-parameter
 models applied to the magnetosphere. The ``linear+noise" model was abandoned because of observed
 nonlinearity, low dimensionality and lack of long-term predictability in the auroral index time series. Low
 dimensional models have been questioned because the low dimensionality is not unique to them
 and because their scaling properties are  not robust against changes in the input parameters.
 An alternative, fractional Brownian motion, which gives low dimensionality and robust scaling
 is unsatifying because it does not lead to an underlying plasma physical description. The newest alternative, 
 SOC, chosen for its robust scaling properties, can be seen as providing both a physical explanation 
 for fBm and also accounting for the bursty nature of transport in the magnetosphere.
 SOC has yet to have its low dimensionality and predictability properties fully defined, but so far they seem to be similar to those of fBm. Thus attention must be focused
 on other means of distinguishing these last two, such as the observed intermittency and avalanching properties.
 Even so, questions about the application of fBm and SOC as models of the magnetosphere's large 
 scale output  (picture A) rather than of its solar wind-driven aspects are raised by the similarity of 
 input and output power spectra and burst distributions. Resolution of these issues is hampered by the narrow bandwidth 
 of even the best available data series,  which for example make it difficult to distinguish between wide-band coloured
 noise and random state changes as the origin of the $f^{-2}$ spectrum of AE (Watkins et al., 1999a).

  This seems to leave four possibilities,  not all of which are mutually exclusive:
  
  i) The global ``SOC"-like properties we have referred to come from outside the magnetosphere, i.e. the magnetosphere
  can be quite well described by a ``weakly nonlinear plus coloured noise" model; weakly nonlinear to give
  the necessary degree of predictability of the output from the input while
  giving long-term unpredictibility, but with a coloured noise
  input from the fBm or SOC nature of the turbulent solar wind causing the scaling  properties. 
  This scenario appears to be compatible with picture C,  the data in Freeman et
  al. (2000), and alternative iii) below.
  
  ii) Some SOC systems (Watkins et al, 1999b; Consolini,  private communication, 1998)
   will destroy the information contained in their input. The scaling observed in their outputs is then
  independent of any present in the input, so any common scaling exponents between
input and output   are either coincidental or evidence of universality in certain 
confined plasma systems.
 
 iii)  Measuring global properties is the wrong thing to do, i.e. SOC is not an aspect of the global
 magnetosphere but  relevant more locally to the magnetotail   (compatible with picture B, and
  alternative i) above). This possibility
   is likely to be illuminated by further studies of SOC as a magnetotail model.
 
 or iv) that another type of model  is required (e.g. Chapman, 1999). \\

It is also important to emphasise that the extent to which SOC is observable, and 
distinguishable from other nonlinear physics paradigms (such as those presently used to
study turbulence) is an important generic question in contemporary physics (including
but going beyond, plasma physics). The diversity and quality of the existing ground-based
and space-based magnetospheric databases provide a key testbed with which these
intrinsically interdisciplinary questions can now be addressed; while ongoing investigations 
in astrophysical and laboratory confinement systems, both in plasma physics and elsewhere will
continue to be applicable to the question of magnetospheric SOC.
 
  The authors wish to acknowledge many enjoyable and valuable discussions
with John Barrow, Ben Carreras, Tom  Chang, Giuseppe Consolini, Patrick Diamond,
Per Helander, Henrik Jensen,  John King, David Newman, Christophe
Rhodes, David Riley, George Rowlands, Jouni Takalo, Sunny Tam, Dave Tetreault, Vadim Uritsky,  Zoltan V\"{o}r\"{o}s
and Dave Willis. NWW and SCC would like to acknowledge the  hospitality of the MIT Center for Space Research where some of this work
  was carried out. SCC acknowledges a PPARC Lecturer fellowship and ROD the
  support of Euratom and the UK DTI.
  
  \newpage
  
  \begin{tabular}{|l|llll|} \hline
1.Model: & 2.Linear     & 3.Low -dimensional & 4.fBm   & 5.SOC \\ 
 & (plus noise)    & nonlinear  & nonlinear & sandpile \\
 \hline
1. Property & & & &  \\ 
 \hline
2.Short term predictable & Yes & Yes & No  & ? \\
3.Long term predictable & Yes & No &  No & ? \\ 
4.Global Scaling  & No & Sometimes & Yes  & Yes \\
5.Scaling Regions & Sometimes & Sometimes & Yes & Yes \\
6.Low G-P Dimension & No & Yes & Sometimes  &  ? \\ 
7. Avalanches & No & ? & ?  &  Yes \\ \hline

\end{tabular}

\begin{table}
\vspace{0.in}
\caption{ Four examples of possible approaches to
understanding magnetospheric time series, adapted from Hastings and Sugihara (1993)}
\end{table}

\newpage




Angelopoulos, V., et al., 1996. Multipoint analysis of a bursty bulk flow
event on April 11, 1985. Journal of Geophysical Research 101, 4967-4989. \\

Angelopoulos, V., Mukai, T., Kokubun, S., 1999. Evidence for intermittency 
in Earth's plasmasheet and implications for self-organised criticality.
Physics of Plasmas 6, 4161-4168. \\

Aschwanden, M. J.,  Dennis, B. R., Benz, A. O., 1998. Logistic avalanche processes, elementary time
structures and frequency distributions in solar flares. Astrophysical Journal 497, 972-993.\\

Bak, P., 1994. Self-organized criticality: consequences for statistics and predictability of earthquakes.
 In: Newman, W. I., Gabrielov, A., Turcotte, D. L., (Eds.), Nonlinear Dynamics and Predictability of Geophysical
Phenomena. American Geophysical Union.\\

Bak, P., 1997. How Nature
Works:  The Science of Self Organised Criticality. Oxford University
Press. \\

Bak, P.,   Tang, C.,   Wiesenfeld, K., 1987. Self--organized criticality: an explanation of
1/f noise. Physical  Review  Letters  50, 381--384.\\

Bak, P., Tang, C., Wiesenfeld, K., 1988.  Self--organized criticality. Physical Review A 38, 364-374.\\

Baker, D. N., Klimas, A. J., McPherron, R. L., Buchner, J., 1990. The evolution from weak to strong
geomagnetic activity: an interpretation in terms of deterministic chaos. Geophysical Research Letters
17, 41-44.\\

Baker, D.N.,  Klimas, A. J., Vassiliadis, D., Pulkkinen, T. I., McPherron, R. L., 1997.
Reexamination of driven and unloading aspects of magnetospheric substorms. Journal of Geophysical Research 102,
7169-7177.\\

Baker, D.N.,  Kanekai, S. G., Klimas, A. J., Vassiliadis, D., Pulkkinen, T. I., 1999.
Collective phenomena in the inner magnetosphere.
Physics of Plasmas 6, 4195-4199.\\

Bargatze, L. F., Baker, D. N., McPherron, R. L., Hones, E. W., 1985.
Magnetospheric impulse response for many levels of geomagnetic activity. 
Journal of Geophysical Research 90, 6387-6394.\\

Bendat, J. S., 1958. Principles and Applications of Random Noise Theory. John Wiley.\\

Boffetta, G., Carbone, V., Giuliani, P., Veltri, P., Vulpiani, A., 1999.
Power laws in solar flares: Self-organised criticality or turbulence?
Physical Review Letters 83, 4662-4665.\\

Borovsky, J. E., and Nemzek, R. J.,  1994. Substorm statistics: occurrences and amplitudes. In: Proceedings of the
Second International Conference on Substorms, Fairbanks, Alaska. Geophysical Institute. \\

Cardy, J. L., 1996. Scaling and Renormalization in Statistical Physics.
 Cambridge University Press. \\

Carreras, B. A., van Milligen, B.,  Hidalgo, C., Balbin, R., Sanchez, E., Garcia-Cortes, I., Pedrosa, M. A.,
Bleuel, J., Endler, M., 1999. Self-similarity properties of the probability
distribution function of turbulence-induced particle fluxes at the plasma
edge. Physical Review Letters 83, 3653-3656.\\

Chang, T.  S., 1992. Low
dimensional behaviour and symmetry breaking of stochastic systems near
criticality - can these effects be observed in space and in the
laboratory ?.  IEEE Transactions on  Plasma Science  20, 691--694.\\

Chang, T.  S., 1999. Self-organized criticality, multi-fractal spectra, sporadic localized
reconnections and intermittent turbulence in the magnetotail. Physics of Plasmas 6, 4137-4145.\\

Chapman, S. C., 1999. A deterministic avalanche model with limit cycle exhibiting period
doubling, intermittency and self-similarity. Submitted to Physical Review E. \\
 
Chapman, S. C., Watkins, N. W., Dendy, R. O., Helander, P.,    Rowlands, G., 1998.
A simple avalanche model as an analogue for magnetospheric activity. Geophysical Research Letters 25, 2397-2400. \\

Chapman, S. C., Dendy, R. O., Rowlands, G., 1999. A sandpile model with dual scaling
regimes for laboratory, space and astrophysical plasmas. Physics of Plasmas 6, 4169-4177. \\

Chapman, S. C., Watkins, N. W., Rowlands, G., 2000. Signatures of dual scaling regimes in a simple avalanche
model for magnetospheric activity. In press, Journal of Atmospheric and Solar-Terrestrial Physics.\\

Chen, J., and Holland, D. L., 1993.  Nonlinear dynamics of particles in the magnetosphere: a tutorial review.
 In: Chang, T. S., and Jasperse, J. R. (Eds.), Physics of Space Plasmas (1993). MIT Center for Theoretical
 Geo/Cosmo Plasma Physics.\\

Consolini, G., Marcucci, M. F., Candidi, M., 1996. Multifractal structure of auroral electrojet
data. Physical Review Letters 76, 4082-4085. \\

Consolini G., 1997. Sandpile cellular automata and magnetospheric dynamics. In:  Aiello, S.,  Iucci, N.,
  Sironi, G., Treves A., Villante, U., (Eds.),
 Cosmic Physics in the Year 2000.  Proc. vol. 58, Societa Italia di Fisica.  \\

Consolini, G., 1999. Avalanches, scaling and 1/f noise in magnetospheric dynamics. Submitted to Physical  Review  Letters.\\

Davis, T. N.,  and  Sugiura, M., 1966.
Auroral electrojet activity index AE and its universal time variations, 
Journal of Geophysical Research 71, 785-801.\\

Dendy, R. O.,  and  Helander, P., 1997. Sandpiles, silos and tokamak phenomenology:  a
brief review. Plasma Physics and Controlled Fusion 39,
1947--1961.\\

Dendy, R. O.,  and  Helander, P., 1998. On the appearance and non--appearance of
self--organised criticality in sandpiles.  Physical  Review  E 57, 3641-3644, 1998.\\

Einaudi, G., and Velli, M., 1999. The distribution of flares, statistics of magnetohydrodynamic 
turbulence and coronal heating. Physics of Plasmas 6, 4146-4153. \\

Freeman, M. P., Watkins, N. W., Rhodes, C., 1998. On the two-component nature of the auroral electrojets. 
AGU Fall Meeting, EOS Transactions, 79 supplement, F793. \\

Freeman, M. P., Watkins, N. W, Riley, D. J.,  2000. Evidence for a solar wind origin of the power law
burst lifetime distribution of the AE indices. 
Geophysical Research Letters 27, 1087-1090.\\

Grassberger, P., and Procaccia, I., 1983. Characterisation of strange attractors.
Physical Review Letters, 50, 346-349.\\

Hastings, H. M., and Sugihara, G., 1993. Fractals: a Users Guide for the Natural Sciences. Oxford
University Press. \\

Horbury, T. S., and Balogh, A., 1997. Structure function measurements of the intermittent MHD turbulent
cascade. Nonlinear Processes in Geophysics, 4, 185-199.\\

Horton, W., and Doxas, I., 1996. A low dimensional energy conserving state space model for substorm
dynamics. Journal of Geophysical Research, 101, 27223-27237. \\

Horton, W., Smith, J. P., Weigel, R., Crabtree, C., Doxas, I., Goode, B., Cary, J., 1999. 
The solar-wind driven magnetosphere-ionosphere as a complex dynamical system.
Physics of Plasmas 6, 4178-4184. \\
  
Huang K., 1987. Statistical Mechanics, Second Edition. John Wiley.\\

Hwa, T., and Kardar, M., 1992. Avalanches, hydrodynamics, and discharge events in
models of sandpiles.  Physical Review A 45, 7002-7023. \\

Jensen, H. J.,  Christensen, K.,  Fogedby, H. C., 1989. 1/f noise, distribution
of lifetimes, and a pile of sand. Physical Review B 40, 7425-7427. \\
 
Jensen, H.  J.,  1998. Self-Organised Criticality:  Emergent Complex Behaviour in Physical
and Biological Systems. Cambridge University Press.\\

Kamide, Y.,  and Baumjohann, W., 1991. Magnetosphere-Ionosphere Coupling. Springer Verlag. \\

Katz, J. I., 1986. A model of propagating brittle failure in
heterogeneous media. Journal of Geophysical
 Research  91, 10412--10420.\\ 

Klimas, A. J., Baker, D. N., Roberts, D. A., Fairfield, D. H., Buchner, J., 1992. A nonlinear
dynamical analogue model of geomagnetic activity. 
Journal of Geophysical Research 97, 
12253-12266.\\
 
Klimas, A. J., Vassiliadis, D., Baker, D. N., Roberts, D. A., 1996.
 The organised nonlinear dynamics of the magnetosphere. Journal of Geophysical
 Research 101, 13089--13113.\\
 
Lichtenberg, A. J.,  and Lieberman, M. A., 1992. Regular and Chaotic Dynamics, Second Edition. Springer Verlag.\\

Lu, E. T., 1995. Avalanches in continuum dissipative systems. Physical Review Letters 74, 2511-2514.\\

Malamud, B. D., and Turcotte, D. L., 1999. Self-Affine Time Series: I. Generation
and Analyses. In: Dmowska, R., and Saltzman, B. (Eds.), Advances in Geophysics, Volume 40,
 Long-Range Persistence in Geophysical Time Series. Academic Press.\\

Osborne, A. R., and Provenzale, A.,1989. Finite correlation dimension for stochastic
systems with power-law spectra. Physica D 35, 357-381. \\

Palmer, T. N., 1993. A nonlinear dynamical perspective on climate change.
Weather 48, 314-326. \\

Pedrosa, M. A., Hidalgo, C., Carreras, B. A., Balbin, R., Garcia-Cortes, I.,
Newman, D., van Milligen, B.,    Sanchez, E., 
Bleuel, J., Endler, M., Davies, S., Matthews, G. F., 1999. Empirical similarity of frequency spectra of the edge-plasma fluctuations
in toroidal magnetic-confinement systems. Physical Review Letters 82, 3621-3624.\\

Pelletier, J. D.,  and Turcotte, D. L., 1999. Self-affine time series: II.  
 applications and models. In: Dmowska, R., and Saltzman, B. (Eds.), Advances in Geophysics, Volume 40,
 Long-range Persistence in Geophysical Time Series. Academic Press.\\

Pinho, S. T. R., and Andrade, R. F. S., 1998. An Abelian model for rainfall. Physica A 255, 483-495. \\

Robinson, P. A., 1993. Self-organized criticality. In: Chang, T. S., and Jasperse, J. R. (Eds.), 
Physics of Space Plasmas (1993). MIT Center for Theoretical
 Geo/Cosmo Plasma Physics. \\ 

Roberts, D. A., and Goldstein, M. L., 1987. Spectral signatures of jumps
and turbulence in interplanetary speed and magnetic field data. Journal of Geophysical Research 92, 
10105-10110.\\

Rodr\'{i}guez-Iturbe and Rinaldo, 1997.  Fractal River
Basins:  Chance and Self-Organisation. Cambridge University Press.\\

Rowlands, G., 1990. Non-Linear Phenomena in Science and Engineering. Ellis Horwood. \\

Sharma, A.  S., 1995. Assessing the magnetosphere's nonlinear behaviour:  its dimension is
low, its predictability high. Reviews of Geophysics 33, Part 1,
Suppl. S., 645--650. \\

Shaw, R.,  1984. The Dripping Faucet as a Model Chaotic System. Aerial Press.\\

Takalo, J., 1993. Correlation dimension of AE data, Ph. Lic. Thesis, Laboratory Report 3. Department of Physics,
University of Jyvaskyla.\\

Takalo, J., Timonen, J., Koskinen, H., 1994. Properties  of AE data and
bicolored noise. Journal of Geophysical Research 99, 13239-13249.\\
 
Takalo, J., Lohikoski, R.,  Timonen, J., Lehtokangas, M., and Kaski, K., 1996. Time series
analysis and prediction of AE and Dst data. In: Burke, W. and Guyenne, T. D., (Eds.),
Environment Modelling for Space Based Applications, ESA SP-392. European Space Agency.\\

Takalo, J., Timonen, J., Klimas, A. J., Valdivia, J. A., Vassiliadis, D., 1999a. Nonlinear energy
dissipation in a cellular automaton magnetotail field model. Geophysical Research Letters 26, 1813-1816.\\

Takalo, J., Timonen, J., Klimas, A. J., Valdivia, J. A., Vassiliadis, D.,1999b.
 A coupled-map model for the magnetotail current sheet. Geophysical Research Letters,
 26, 2913-2916.\\

Takalo, J., Timonen, J., Klimas, A. J., Valdivia, J. A., Vassiliadis, D., 1999c. A coupled-map
as a model of the dynamics of the magnetotail current sheet. Submitted to
 Journal of Atmospheric and Solar-Terrestrial Physics. \\

Tsurutani, B.,  Sugiura, M.,  Iyemori, T.,   Goldstein, B. E., 
Gonzalez, W. D.,   Akosofu, S.-I.,  Smith, E. J., 1990. The nonlinear response of
AE to the IMF $B_s$ driver:  a spectral break at 5 hours. Geophysical Research Letters 17, 279-282.
\\

Uritsky, V.  M., and  Pudovkin, M. I., 1998. Low frequency 1/f-like
fluctuations of the AE-index as a possible manifestation of
self-organised criticality in the magnetosphere.  Annales  Geophysicae 16, 1580-1588.\\

Vassiliadis, D., Anastasiadis, A., Georgioulis, M., Vhlahos, L., 1998.
Derivation of solar flare cellular automaton models from a subset of the magnetohydrodynamic equations.
Astrophysical Journal  509, L53-56. \\

V\"{o}r\"{o}s, Z., 1991. Synergetic approach to substorm phenomenon. In: Kan, J. R., Potemra, T. A.,
Kokubun, S., Iijima, T.,  (Eds.), Magnetospheric Substorms. American Geophysical Union.\\

V\"{o}r\"{o}s, Z.,  Kovacs, P., Juhasz, A., Kormendi, A., Green, A. W., 1998. Scaling
laws from geomagnetic time series. Geophysical Research Letters 25, 2621-2624.\\

Watkins, N. W., Freeman, M. P., Rhodes, C. S., Rowlands, G. 1999a. Ambiguities
in determination of self-affinity in ionospheric time
series.  Submitted to proceedings of the EGS/AGU Richardson
Conference, Roscoff, France, June 1998.\\

Watkins, N. W., Chapman, S. C., Dendy, R. O., Rowlands, G., 1999b. Robustness
of collective behaviour in a strongly driven avalanche model: magnetospheric
implications. Geophysical Research Letters 26, 2617-2620.\\

Zapperi, S., Lauritsen, K. B., Stanley, H. E., 1996. Self-organized branching
processes: a mean field theory for avalanches. Physical Review Letters 75,
4071-4074.\\
 
Zelenyi, L. M., Milovanov, A. V., Zimbardo G., 1998. Multiscale magnetic structure of the distant
tail: self-consistent fractal approach. In: Nishida, A., Baker, D. N., Cowley, S. W. H. (Eds.),
New perspectives on the earth's magnetotail. American Geophysical Union.\\
 
\newpage

\end{document}